# Observation of reduced 1/f noise in Graphene field effect transistors on Boron Nitride substrates

Morteza Kayyalha[1,2], Yong P. Chen[1,2,3,*]

[1] Birck Nanotechnology Center, Purdue University, West Lafayette, IN 47907

[2] School of Electrical and Computer Engineering, Purdue University, West Lafayette, IN 47907

[3] Department of Physics and Astronomy, Purdue University, West Lafayette, IN 47907

[*] To whom correspondence should be addressed: yongchen@purdue.edu

**Abstract**

We have investigated the low frequency (f) flicker (also called 1/f) noise of single-layer graphene devices on h-BN (placed on $SiO_2$/Si) along with those on $SiO_2$/Si. We observe that the devices fabricated on h-BN have on average one order of magnitude lower noise amplitude compared with devices fabricated on $SiO_2$/Si. We associate this noise reduction to the lower densities of impurities and trap sites in h-BN than in $SiO_2$. Furthermore, the gate voltage dependent noise amplitude shows a broad maximum at Dirac point for devices on h-BN, in contrast to the M-shaped behavior showing a minimum at Dirac point for devices on $SiO_2$, consistent with the reduced charge inhomogeneity (puddles) for graphene on h-BN. This study demonstrates that the use of h-BN as a substrate or dielectric can be a simple and efficient noise reduction technique valuable for electronic applications of graphene and other nanomaterials.

**Introduction**

Due to its excellent transport properties and potential in future nano-electronics, graphene has attracted a lot of attention over the past decade [1]–[5]. However, lack of sufficient energy gap in graphene limits its applications in digital transistors and electronics. On the other hand, it is shown that graphene may be highly promising for applications in analogue electronics, radio frequency (RF) devices, and sensors where high mobility, gate tunability, and sensitivity to external impurities are of paramount importance [3], [6]–[8]. In these applications, low frequency flicker noise also plays a significant role in the performance of the nanoscale devices; particularly because the noise amplitude generally increases as device dimensions shrink [9]–[17].

Several experimental studies have investigated the flicker or 1/f noise performance of single-layer or multi-layer graphene devices fabricated on $SiO_2$ substrates [9]–[12], [14]–[16], [18]–[20] and on SiC substrates [21]. It is generally believed that flicker noise originates from the fluctuations in the number of charge carriers in the channel due to trapping/detrapping of carriers in the oxide layer and/or fluctuations in the mobility of carriers [9], [22], [23]. Also, it has been shown that the carrier mobility in graphene can be highly sensitive to the substrate material, with significant enhancement in mobility at low temperatures when $SiO_2$ is replaced with h-BN [24]–[26]. Therefore, a study of low frequency noise in graphene transistors with h-BN as the substrate is desired for a more complete understanding of the device performance of graphene on h-BN. Such a study is also of interests for the nano-electronics community, where different techniques are actively being explored to reduce the low frequency noise of new materials and devices.

In this work, we investigate the low frequency noise performance of single-layer graphene field effect transistors on h-BN substrates and compare it to those on $SiO_2$ substrates (standard doped Si wafer with 300 nm $SiO_2$), at otherwise similar conditions. Even though the measured *room temperature* mobility does not show significant differences between devices on h-BN and $SiO_2$ substrates, the normalized noise amplitude is on average one order of magnitude lower for devices on h-BN substrates. We believe that this reduction in the 1/f noise amplitude is due to the fact that h-BN has less charge impurities and trapping sites compared to $SiO_2$ [27] that may help reduce the fluctuations in both mobility and the numbers of carriers in the channel.

**Device Fabrication and Methods**

We prepare h-BN flakes (with thickness ranging from 20 to 50 nm) using the standard scotch tape exfoliation technique [2], [24], [28], [29] from commercial h-BN (Momentive Performance Materials Inc.). We then transfer the h-BN flakes onto a substrate of 300 nm-thick $SiO_2$ on highly doped Si substrate (Nova Electronic Materials). The doped Si will be used as the back gate for our devices. Single-layer graphene flakes are exfoliated also using the scotch tap technique from high quality polycrystalline graphite (Momentive Performance Materials Inc.) onto a polymer film consisting of a thin layer of positive resist (polymethyl methacrylate, PMMA) on top of a thin layer of polyvinyl alchohol (PVA). Using a homemade transfer stage, the graphene flakes are then transferred either on top of $SiO_2$ or h-BN substrates. The single layer thickness of graphene is confirmed by its characteristic optical contrast and Raman spectrum (measured by a Horiba XploRA Raman microscope with 532 nm excitation laser, with an example shown in Fig. 1b) [30]–[32]. Finally e-beam lithography is used to design the contact patterns, followed by deposition of Cr (10 nm) and Au (60 nm) contact electrodes. Figure 1a shows an optical image of a representative device on top of an h-BN substrate. We note that no significant change in the noise performance of graphene devices on $SiO_2$ is observed when graphene is prepared using the standard exfoliation and transfer process [2], [28], [29] without the use of PMMA/PVA films.

Our noise measurements were performed under ambient conditions (room temperature and atmospheric pressure) using a two-probe measurement technique depicted in Figure 1c. In this technique, a low noise preamplifier (SR560) provides a "silent" (low noise) constant DC voltage bias ($V_{DS}$ = 40 mV for all presented data here unless stated otherwise) across the source and drain of the transistor and the same amplifier is utilized to amplify the signal (and noise) of the source-drain current (I). The spectral density of the current noise ($S_I$) is then monitored using a dynamic signal analyzer (Agilent 35670A Dynamic Signal Analyzer). The back-gate voltage ($V_g$) is supplied by a Keithley 2400 source-meter.

The spectral density ($S_I$) of the Flicker noise (also known as 1/f noise, where f is the frequency) in the current (I) can be expressed as $S_I = \frac{AI^2}{f^\beta}$, where A is the (dimensionless) noise amplitude, and β is the frequency exponent with a value around 1 [9], [22]. We observe that the noise spectral density of our devices, regardless of their substrates, is proportional to $I^2$ (varied by varying $V_{DS}$ at fixed $V_g$) as previously reported for graphene on $SiO_2$ [9], [14], [18], [33]. Normalized spectral densities of the measured noise signal ($S_I/I^2$) on a representative graphene device on 40 nm h-BN vs. frequency (f) and at three different $V_g$'s are plotted in Figure 1d. As it can be seen, $S_I/I^2$ is proportional to $1/f^\beta$ with β ranging from 0.85 to 1.2 for our device.

**Measurements and discussion**

Figure 2a shows four-probe resistances (R) of two graphene transistors on $SiO_2$ and h-BN substrates respectively vs. the back-gate voltage ($V_g$-$V_D$, with respect to the Dirac voltage, $V_D$). Using the Drude's formula ($\mu_{FE} = \frac{1}{C_g}\frac{d\sigma}{dV_g}$, where $C_g$ is the gate capacitance per unit area and σ is the four-probe conductivity of the channel material), we obtain field effect mobility $\mu_{FE}$ ~ 4000-6000 cm$^2$/V-s for both devices. Figure 2b shows the drain current (I) vs. $V_g$-$V_D$ of the same devices of Figure 2a. These measurements are also done at room temperature and immediately before the noise measurements. Measured drain current values here are later used to calculate the normalized noise spectral density ($S_I/I^2$) as well as the noise amplitude (A).

In order to have a better comparison between the noise behavior of graphene devices on top of different substrates, we calculate the noise amplitude as $A = \frac{1}{n}\sum_{i=1}^{n}\frac{f_i S_{Ii}}{I^2}$, which is a noise characteristic independent of the current passing through the transistor and averaged over n different frequencies. The quantity (A) and also the normalized noise amplitude (A×W×L, where W and L are the width and length of the channel, respectively) will be used as metrics to indicate how much improvement in noise we gain by changing different parameters in our devices (e.g. changing the substrate from $SiO_2$ to h-BN), with larger A indicating a worse noise performance.

Figure 2c and d show the noise amplitude (A) and the normalized noise amplitude (A×W×L) vs. $V_g$-$V_D$ in devices studied in Figures 2a and b at room temperature. We notice two significant differences between the graphene device on $SiO_2$ and that on h-BN. First, the normalized noise amplitude is around one order of magnitude lower in the graphene device on h-BN compared to that on $SiO_2$. Previous experiments have demonstrated that h-BN substrate has much lower densities of charge impurities and trap sites compared with $SiO_2$, leading to significantly increased mobility of graphene at low temperatures (where impurity scattering is especially important for limiting the mobility) [24]–[26]. Since the flicker noise originates from fluctuations in the number or mobility of carriers, the reduced impurities in the h-BN is expected to help reduce these fluctuations and result in lower noise. Secondly, we observe an M-shaped behavior (with a *minimum* at Dirac point) in the noise amplitude near the Dirac point in the graphene device on $SiO_2$, consistent with previous experiments [12]. This behavior is attributed to the existence of electron-hole puddles and charge inhomogeneity near the Dirac point. In contrast, such an M-shaped behavior is not observed for the device on h-BN, where the measured noise amplitude vs. $V_g$ shows a qualitatively similar trend as R vs. $V_g$ and a *maximum* around the Dirac point (where the channel resistance is the largest). The absence of the M-shaped behavior in graphene on h-BN (which we confirmed in multiple devices) is consistent with the expectation that the h-BN substrate can substantially reduce the charge inhomogeneity and electron hole puddles in graphene [24], [34]. The observed gate-dependence of the noise amplitude in graphene devices on h-BN is qualitatively consistent with Hooge's empirical relation (A = $\alpha_H$/N, where N is the total number of carriers, which is tuned by the gate, and $\alpha_H$ is the Hooge's noise parameter) [9], [33].

Figure 3 plots a histogram of the normalized noise amplitudes measured at all the gate voltages for all graphene devices studied, both on h-BN (data in blue, from 6 devices) and $SiO_2$ (data in red, from 3 devices; we also included 3 devices from a previous study [12], shown as data in black). We observe that the noise amplitude of graphene devices is reduced on average by one order of magnitude with h-BN substrate vs. $SiO_2$ substrate.

**Conclusion**

We have measured the current noise spectral density of graphene field effect transistors on two different substrates, h-BN and $SiO_2$. We have observed that the normalized noise amplitude of graphene transistors measured at room temperature is reduced by ~10 times for devices on h-BN compared to devices on $SiO_2$. Furthermore, the gate voltage dependence of the noise amplitude exhibits qualitatively different behaviors for devices on h-BN (with a noise amplitude maximum at Dirac point) versus those on $SiO_2$ (with a noise amplitude minimum at Dirac point and an M-shaped gate dependence). Our observations can be attributed to the significantly reduced charge impurities and traps in h-BN compared to $SiO_2$, leading to significantly reduced fluctuations in the carrier mobility and density (including charge puddles) in graphene. Our work demonstrates that the use of h-BN substrates can offer an efficient technique to reduce the noise and improve the performance for graphene-based electronic devices. This approach may also be extended to electronic devices based on other nanomaterials.

**Note**

After our manuscript was submitted, we became aware of the publication of a related work [35].

**Acknowledgement**

We thank School of electrical and computer engineering of Purdue University for support.

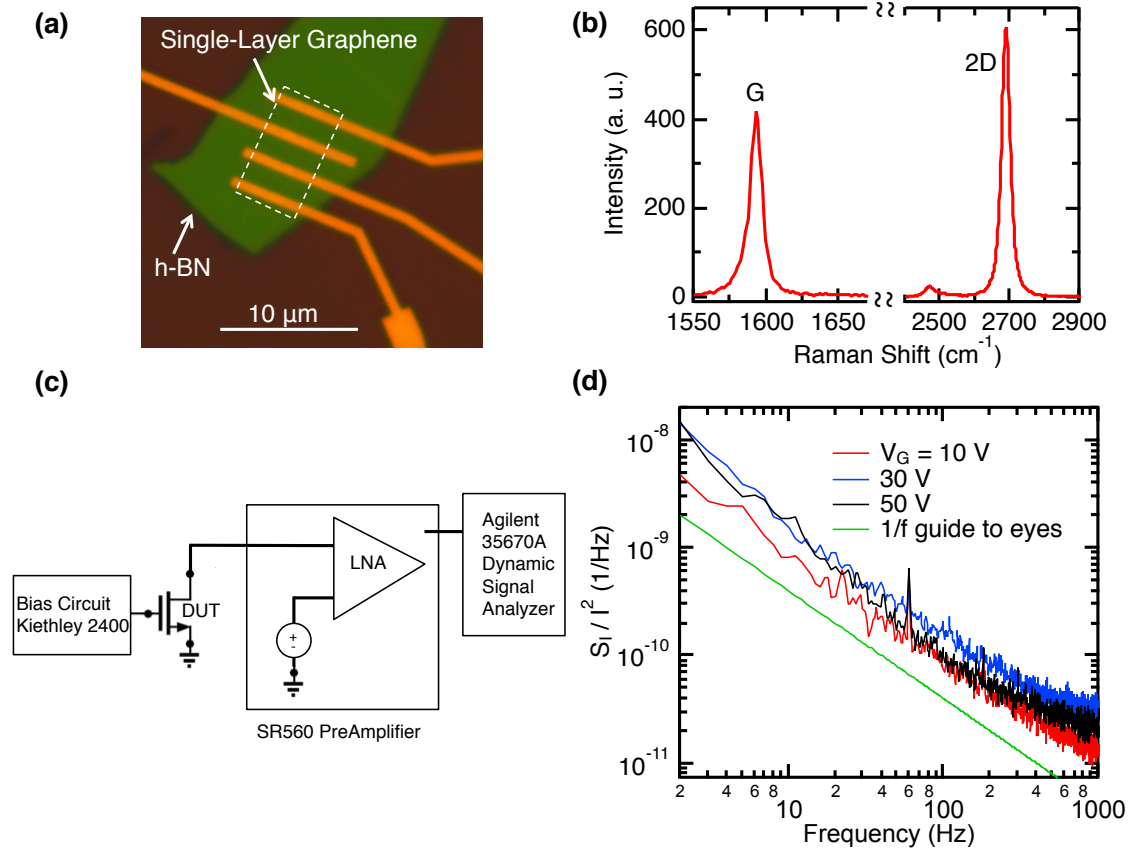

Figure 1. (a) Optical microscope image of a back-gated field effect transistor based on single-layer graphene on h-BN (on a $SiO_2$/Si substrate). Scale bar is 10 μm. (b) Raman spectrum (excitation laser wavelength = 532nm) of a single-layer graphene on h-BN. The 2D peak at 2680 cm$^{-1}$ has higher intensity than G-peak and has a single-Lorentzian shape with full-width-at-half-maximum (FWHM) ~ 30 cm$^{-1}$, indicating a single-layer graphene. (c) Schematic of a two-probe noise measurement set-up used in our experiment. A preamplifier (SR560) is used to bias the device and also amplify the current signal that contains the noise. The spectral density of the current noise is measured with a dynamic signal analyzer. (d) Normalized spectral current noise density ($S_I/I^2$) of the device shown in (a) as a function of frequency (f) at various back-gate voltages ($V_g$). Green line (slope=-1 on this log-log plot) is a guide to eyes for the 1/f behavior.

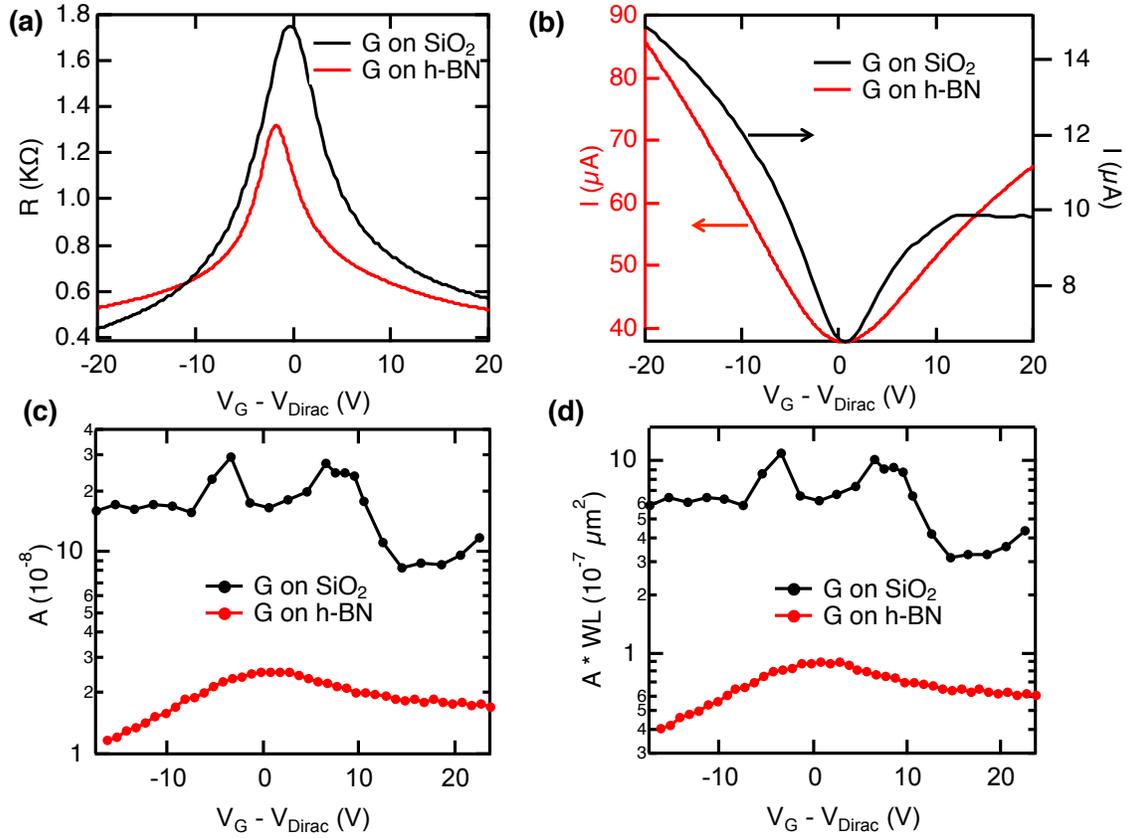

Figure 2. (a) Four-probe resistance (R) of the graphene channel measured using an AC technique with an SR830 lock-in amplifier vs. $V_g-V_D$ (gate voltage with respect to the Dirac point voltage) for a device on h-BN (red curve) substrate and another on $SiO_2$ (black curve). (b) DC drain current (I) vs. $V_g-V_D$ of the same devices in (a) measured right before the noise measurements. (c) Amplitude of the noise (A) vs. $V_g-V_D$ and (d) normalized noise amplitude (A*WL) vs. $V_g-V_D$ for the devices presented in (a).

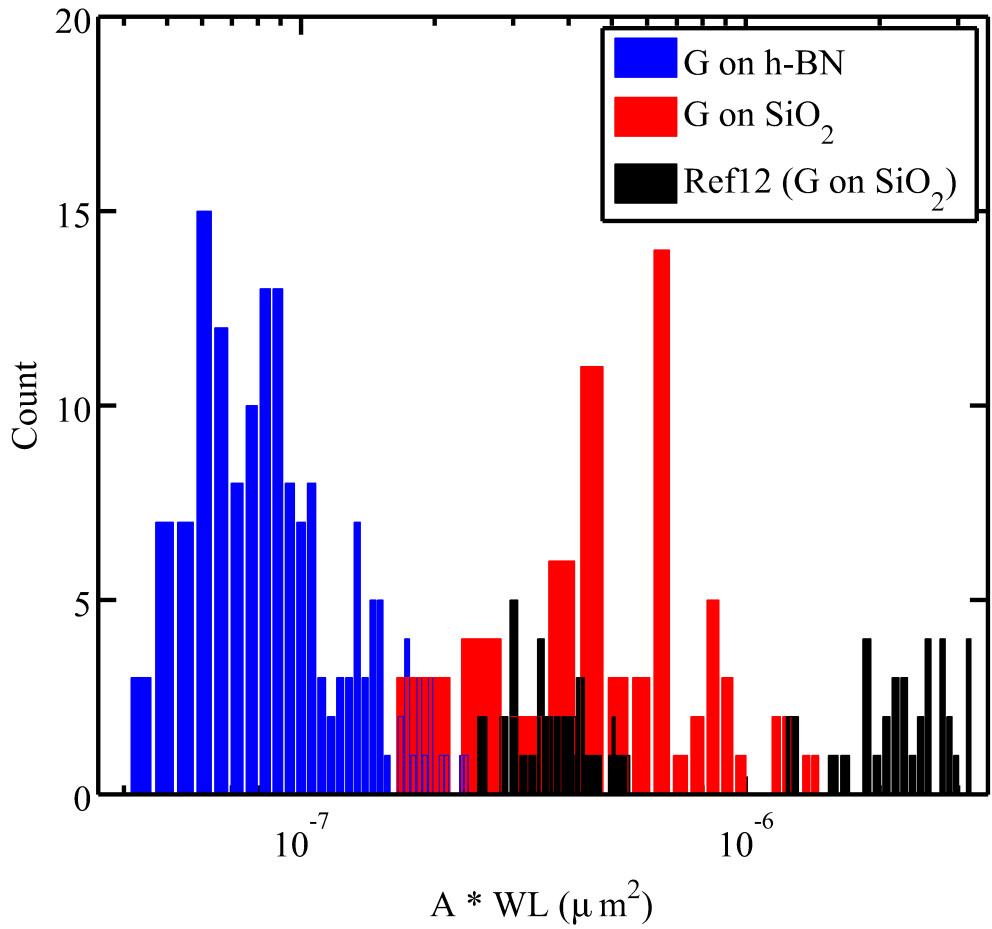

Figure 3. Histogram of the normalized noise amplitude (A*WL) at all the gate voltages measured in all graphene transistors on h-BN (blue, including 6 devices) and on $SiO_2$ (red, including 3 devices and black, including 3 devices of [12]) substrates.